# DESIGN OF NEW HELIUM VESSEL AND TUNER FOR CEPC 650 MHZ 2-CELL CAVITY*

Z. H. Mi[†], Z. Q. Li, P. Sha, J. Y. Zhai, F. S. He, Q. Ma, B. Q. Liu, X. Y. Zhang, R. X. Han,
F. B. Meng, H. J. Zheng, Key Laboratory of Particle Acceleration Physics and Technology,
Institute of High Energy Physics, CAS, Beijing, China

## Abstract

CEPC will use 650 MHz cavities for the collider. Each collider cryomodule contains six 650 MHz 2-cell cavities, which is totally new. Therefore, new helium vessel and tuner are designed for the 650 MHz 2-cell cavity. Also, a test cryomodule, which consists of two 650 MHz 2-cell cavities, has begun as the first step to the full-scale cryomodule. This paper mainly focuses on the structure design of Helium vessel and tuner for the 2-cell cavity.

## INTRODUCTION

Baseline layout and parameters for CEPC Ring SRF system have been public [1]. IHEP is developing 2-cell elliptical 650MHz cavities for CEPC. There're 240 650 MHz 2-cell cavities in total. Each cavity with one set helium vessel and tuner. There 2-cell cavity system are somewhat different from other 650 MHz cavities. The cavity string will be mounted inside the cryostat by means of bottom support and the cavity two beam pipe both with HOM coupler installed lead to the space is tight. For the convenience of cavity installation and processing a new type of helium vessel is developed, which uses square flange end plate. Also, new double lever tuner intended for the cavities has been developed for final tuning of the resonance frequency of the cavity after cooling down and to compensate the resonance frequency variations of the cavity during operation coming from liquid helium pressure fluctuations. The design of helium vessel and tuner have been completed, and the welding sequence of helium vessel is established. The design results of helium vessel and tuner are presented.

## HELIUM VESSEL DESIGN

The helium vessel assembly is constructed form a single 5 mm thick sheet of grade 2 titanium that is rolled and seam welded into a tube with an inside diameter of 450 mm and a length of 382.5 mm. One helium fill port at the bottom of tube for cool-down of the cavity. The helium vessel tube assembly is shown in Fig. 1. It is important to note that the 5 mm wall thickness for the helium vessel tube specification was not to stiffen the cavity, but rather for the effective stroke of tuner acting on the cavity. The df/dp has been reduced to nearly zero during the optimizing of bare cavity. The internal diameter of 2- phase connection pipe is 71 mm. [2-3]

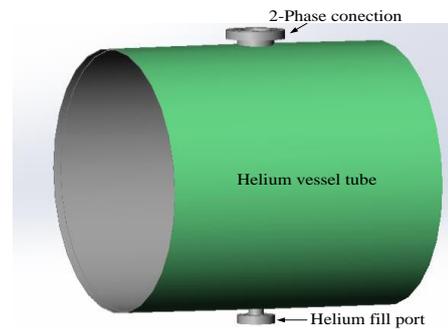

Figure 1: Helium vessel tube assembly.

The endplates of the helium vessel and stiffener are special as shown in Fig. 2.

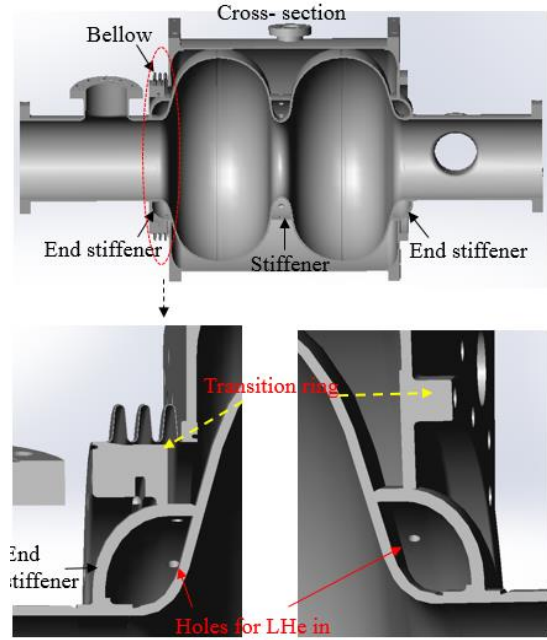

Figure 2: Cross- section of helium vessel with cavity and endplates of helium vessel.

The endplate of the helium vessel at the pickup end with tuning bellows, while the at the coupler end the endplate is fixed. The end stiffeners as parts of the helium vessel. The stiffener with small holes for the liquid helium in. Transition ring with screw holes welding with the stiffener. External tooling will be fixed with the screw holes on the transition ring, during operation 650 MHz cavity. The material for the transition ring is Niobium-Titanium. The main coupler end of the vessel is considered a fixed position relative to the main coupler port of the cavity, while the pickup end a sliding joint design is required to allow for variations in the lengths of the cavity.[4]

___________________
*This study was supported by YOUTH INNOVATION PROMOTION ASSOCIATION CAS.
#mizh@ihep.ac.cn

## WELDING SEQUENCE OF CAVITY WITH HELIUM VESSEL

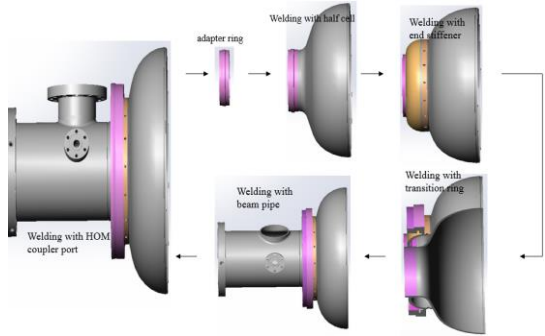

Figure 3: Welding sequence of FP end-group.

As shown in Fig. 3 is the welding sequence of FP end-group. Firstly welding the adapter ring and half-cell together with EBW, secondly welding the end stiffener with adapter ring and half-cell, thirdly welding the transition ring, fourthly welding the beam pipe, fifthly welding the HOM coupler port and FP port.

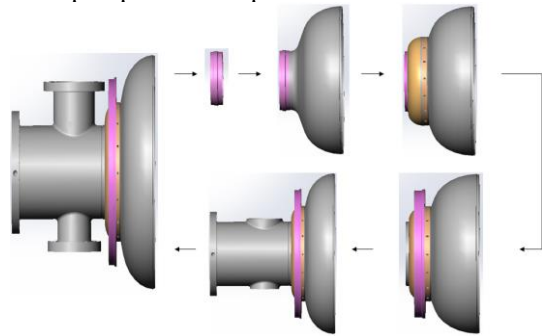

Figure 4: Welding sequence of MC end-group.

Figure 4 is the welding of MC end-group, which welding sequence is same as the sequence of FP end-group welding.

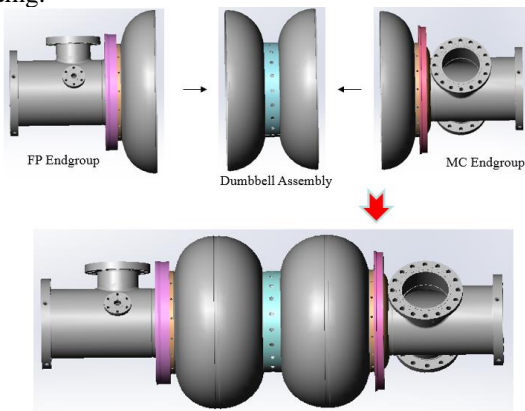

Figure 5: Welding end- group and dumbbell together.

As shown in Fig. 5, welding the FP end-group, MC end-group and dumbly assembly, the next helium vessel tube and endplate will be welding with the cavity, which welding sequence is shown in Fig. 6. Firstly welding the MC end transition ring with square endplate, secondly inert the tube to the cavity welding with the square endplate of the MC end, thirdly TIG welding the tuning bellows with square endplate and transition ring of FP end together, fourthly keep the bellows freely and welding the square endplate with vessel tube.

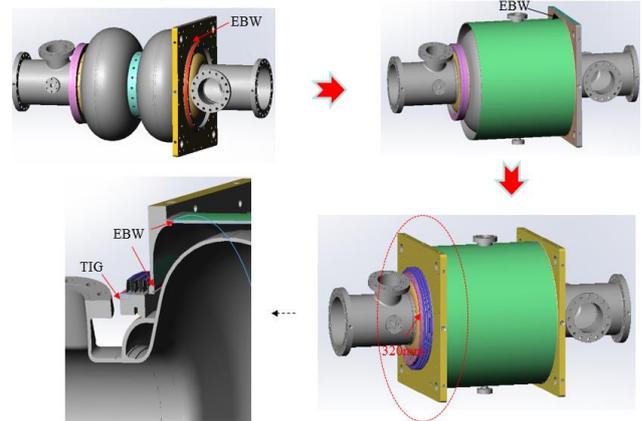

Figure 6: The welding helium vessel with cavity.

## TUNER SYSTEM DESIGN

Due to the space for the tuner is tightly, so a compact fast/slow tuner for the 650 MHz 2-cell cavities has been developed for final tuning of the resonance frequency of the cavity after cooling down and to compensate frequency detuning due to the microphonics. The tuner is improved from the saclay type tuner.

As shown in Fig. 7, one end of the cavity is equipped with a double lever tuner. The lever tuner installed at the PF end, which fixed on the square endplate through two support pedestals. The tuner will take up some straight line space.

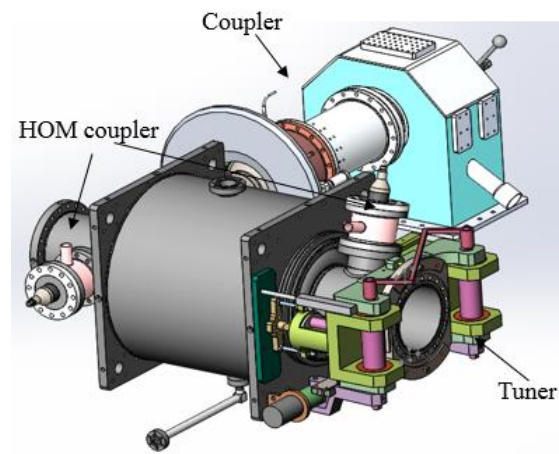

Figure 7: 3D model of cavity + tuner + coupler.

The main parameters of the tuner as shown in Table 1. The slow tuning range is 340 kHz, the fast tuning range is larger than 1.5 kHz. The tuner able to deliver up to 18 kN forces on the cavity.

Table 1: Main Parameters of Double Lever Tuner

| Parameters | Value | Units |
|---|---|---|
| Tuning sensitivity | 310 | kHz/mm |
| Spring Constant | 16 | kN/mm |
| Operating Pressure | ＜5E-5 | Torr |
| Operating lifetime | 20 | Year |
| Coarse (slow) tuner frequency range | 340 | kHz |
| Coarse tuner frequency resolution | ＜20 | Hz |
| Fine (fast) tuner frequency range | ＞1.5 | kHz |
| Fine tuner frequency resolution | 3 | Hz |
| Motor and Piezo temperature | 5~10 | K |
| Motor number | 1 | -- |
| Piezo number | 2 | -- |

As Fig. 8 shown is the working principle of double lever tuner (slow tuner). The motor drives the screw to rotate, causing the lever arms to move toward each other. The lever arms drive the eccentric shaft rotate, causing the tuning arm to move along the axial, and the displacement acting on the flange of beam tube of the cavity. Thus causing the SRF cavity to produce axial deformation.[5-6]

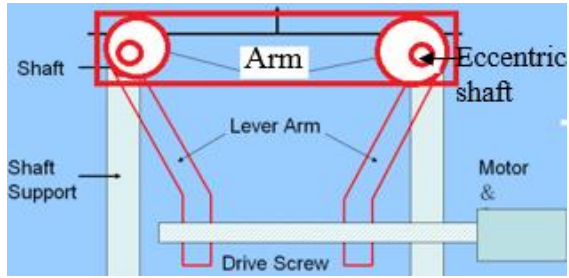

Figure 8: The working principle of double lever tuner.

The tuner 3D model is shown in Fig. 9. Coarse tuner ration larger than 1:10. Tow Piezo installed at the same side of the tuner. For the safety during operation limit switches, limit screws and displacement sensor are installed on the tuner. In order to reduce the quality of the tuner, TA3 is used for the tuner material.

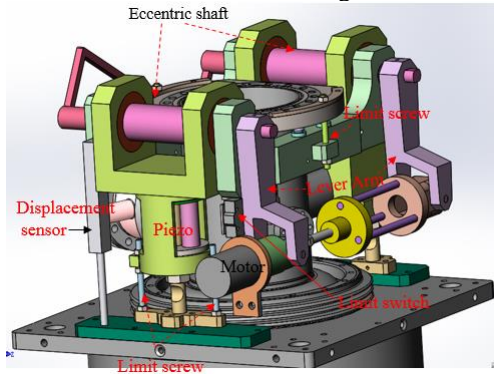

Figure 9: 3D model of double lever tuner.

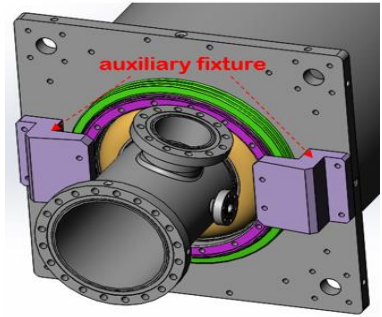

Figure 10: Auxiliary fixture of tuner.

As figure 10 shown, restrained brackets are designed in order to protect cavity during all steps, which can make sure the cavity will be always in elastic region. The auxiliary fixture can be removed after cavity string out of the cleanroom, tuner and limit screw will be installed.

## CONCLUSION

New helium vessel and tuner have been designed for the CEPC 650 MHz 2-cell cavity. The performance of helium vessel and tuner system will be validated in the test cryomodule with two cavities. The test cryomodule may be completed in 2019[7]. Next the overall performance of the components will be evaluated and do further optimal.

## ACKNOWLEDGEMENT

Thanks for the colleagues of IHEP!